\documentclass[
aps,%
12pt,%
final,%
notitlepage,%
oneside,%
onecolumn,%
nobibnotes,%
nofootinbib,%
superscriptaddress,%
noshowpacs,%
centertags]%
{revtex4}

\begin{document}
\selectlanguage{english}

\title{Reconstruction of angular correlations in the associated top quark and the dark matter mediator production}
\author{\firstname{E.}~\surname{Abasov}}
\email{emil@abasov.ru}
\affiliation{Skobeltsyn Institute of Nuclear Physics of Lomonosov Moscow State University (SINP MSU), 1(2), Leninskie gory, GSP-1, Moscow 119991, Russian Federation}
\author{\firstname{L.}~\surname{Dudko}}
\affiliation{Skobeltsyn Institute of Nuclear Physics of Lomonosov Moscow State University (SINP MSU), 1(2), Leninskie gory, GSP-1, Moscow 119991, Russian Federation}
\author{\firstname{E.}~\surname{Iudin}}
\affiliation{Skobeltsyn Institute of Nuclear Physics of Lomonosov Moscow State University (SINP MSU), 1(2), Leninskie gory, GSP-1, Moscow 119991, Russian Federation}
\author{\firstname{A.}~\surname{Markina}}
\affiliation{Skobeltsyn Institute of Nuclear Physics of Lomonosov Moscow State University (SINP MSU), 1(2), Leninskie gory, GSP-1, Moscow 119991, Russian Federation}
\author{\firstname{P.}~\surname{Volkov}}
\affiliation{Skobeltsyn Institute of Nuclear Physics of Lomonosov Moscow State University (SINP MSU), 1(2), Leninskie gory, GSP-1, Moscow 119991, Russian Federation}
\author{\firstname{G.}~\surname{Vorotnikov}}
\affiliation{Skobeltsyn Institute of Nuclear Physics of Lomonosov Moscow State University (SINP MSU), 1(2), Leninskie gory, GSP-1, Moscow 119991, Russian Federation}
\author{\firstname{M.}~\surname{Perfilov}}
\affiliation{Skobeltsyn Institute of Nuclear Physics of Lomonosov Moscow State University (SINP MSU), 1(2), Leninskie gory, GSP-1, Moscow 119991, Russian Federation}
\author{\firstname{A.}~\surname{Zaborenko}}
\affiliation{Skobeltsyn Institute of Nuclear Physics of Lomonosov Moscow State University (SINP MSU), 1(2), Leninskie gory, GSP-1, Moscow 119991, Russian Federation}

\begin{abstract}
For the process of single top quark production within the "simplified model" with a scalar dark matter mediator, a new variable based on angular correlations was presented, for the proper reconstruction of which it is necessary to separate the contributions of two undetectable particles: the neutrino and the mediator. In this work, various machine learning approaches for reconstructing the momenta of these particles are analyzed. A comparison is made between the results obtained using a multilayer perceptron and the Normalizing Flows architectures. The neural networks based on Normalizing Flows, presented in this work, demonstrate a high quality of reconstruction of the target variable and can be used for collider data analysis.
\end{abstract}
\maketitle

\section{Introduction}
\label{sec:intro}
Convincing astrophysical and cosmological observations, such as the studies of galaxy rotation curves~\cite{Rubin:1970zza} and gravitational lensing during galaxy cluster collisions~\cite{Harvey:2015hha}, indicate the presence of a large amount of hidden matter in the Universe, the so-called "dark matter" (DM), which manifests itself through gravitational interactions and interacts weakly with the Standard Model (SM) fields. Based on the assumption that DM is particle in nature, various extensions of the SM are developed to explain the origin of dark matter and its possible detection in experiments.

A potential interaction between DM particles and SM particles may be realized through an intermediate particle, known as the mediator. Depending on the considered mechanism of interaction between DM and SM particles, the mediator is endowed with different properties. In the case of a hypothetically light mediator, models that explicitly include the mediator in the consideration are necessary. These models, known as simplified models~\cite{PhysRevD.79.075020, LHCNewPhysicsWorkingGroup:2011mji}, include particles and interactions beyond the SM. Such simplified models~\cite{Abercrombie:2015wmb,Abdallah:2015ter} can be used in the experimental search for DM particles at colliders. Although there are many different variants of constructing simplified models, those in which DM particles interact with SM particles via new scalar, pseudoscalar, or vector mediators currently appear theoretically attractive.

Assuming minimal flavor violation~\cite{Chivukula1987CompositetechnicolorSM,PhysRevLett.65.2939,BURAS2001161,DAMBROSIO2002155}, third-generation quarks can play a significant role~\cite{PhysRevD.88.063510} in such interactions. Therefore, the study of DM production processes at colliders in association with a top quark is of particular interest, as it has the highest mass. This idea has motivated experimental searches for events in which DM particles are produced in association with a single top quark ($t/\Bar{t}$ + DM) and a top quark pair ($t\Bar{t}$ + DM)~\cite{CMS:2019zzl,ATLAS:2022ygn}. In experimental analyses at colliders, no significant excess above SM predictions has been observed, and the question of increasing analysis efficiency remains relevant.

\begin{figure}[p]
    \centering
    \includegraphics[width=.8\linewidth]{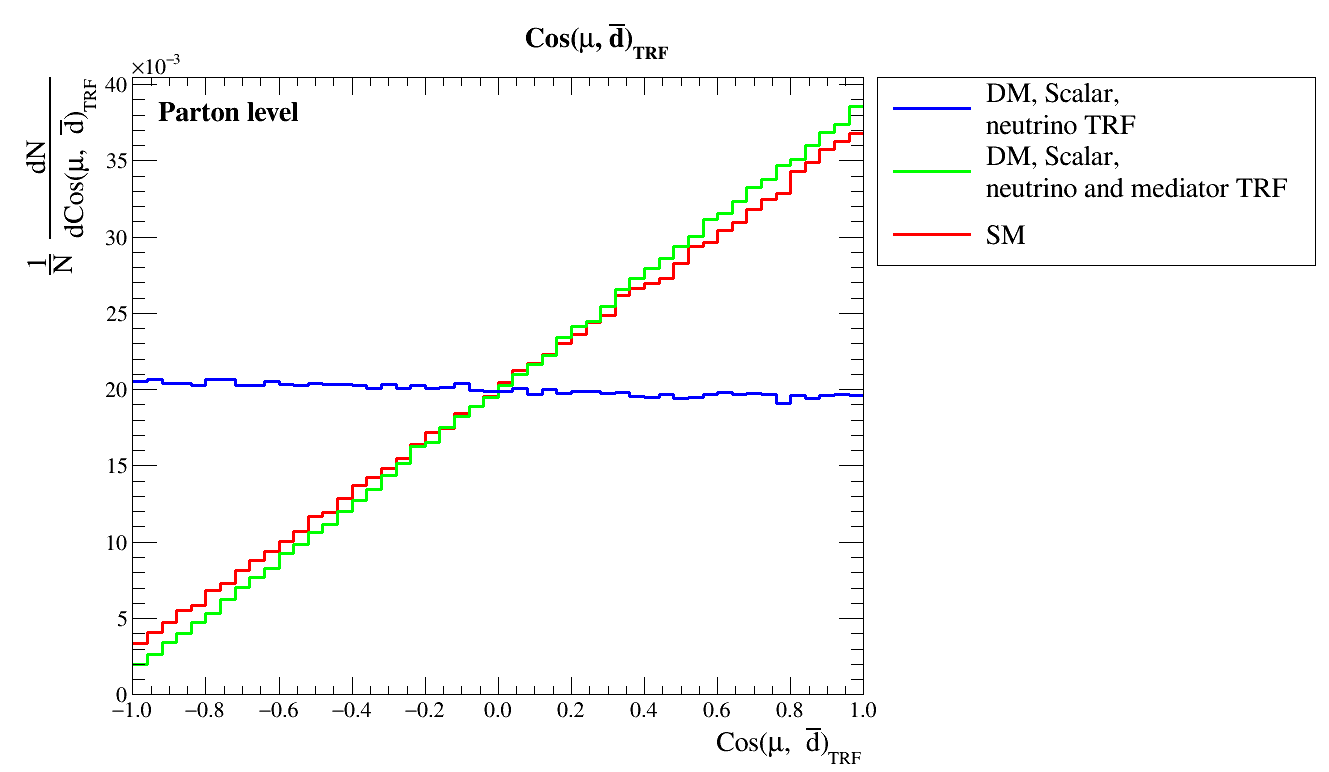}
    \caption{Distribution of the cosine of the angle between the lepton and the down-type quark in the top quark rest frame for the processes $pp\rightarrow t(\rightarrow \nu_l\Bar{l}b)q\ (SM)$ and $pp\rightarrow t(\rightarrow \nu_l\Bar{l}b)q\Bar{\chi}\chi\ $ (neutrino TRF) with the presence of a scalar mediator in the model with $m_\chi = 1$ GeV$,\ m_\phi =400$ GeV $,\ g_\chi=g_\nu=1$. For comparison, the same distribution is shown for the top quark rest frame system that takes the DM mediator into account (neutrino and mediator TRF).}
    \label{fig:spincorrs_parton_scal}
\end{figure}

During the study of single top quark production processes, a new variable was obtained based on the spin correlations of the top quark that separates SM and DM processes~\cite{Abasov:2024nec} (see Fig.~\ref{fig:spincorrs_parton_scal}). This variable can significantly improve the separation between DM and SM processes when it is correctly reconstructed in real events; however, this is accompanied by challenges in reconstructing the top quark rest frame in the presence of two unregistered particles. Proper restoration of the mediator's and neutrino's momenta, and accordingly that variable, is the main task of this work.

\section{Reconstruction of the spin correlations of the top-quark}
\subsection{Data}
\label{sec:data}
At this stage, the analysis is phenomenological; real collider data is not used. For modeling and numerical calculations at the parton level, the computational packages CompHEP4.6rc1~\cite{CompHEP:2004qpa,Pukhov:1999gg} and MadGraph5~\cite{Alwall:2014hca} were employed. 

All calculations are presented for proton-proton collisions at 13 TeV, with the parton distribution functions taken as $\mathrm{NNPDF23\_nlo\_as\_0118}$ from the LHAPDF6 package~\cite{Buckley:2014ana}. In accordance with the recommendations of the LHC Dark Matter Working Group~\cite{Boveia:2016mrp}, the coupling parameter values are set as $g_f=g_\chi=1$, the DM particle mass $m_\chi=1$ GeV, and the DM mediator mass $m_\Phi=400$ GeV, which corresponds to current constraints~\cite{CMS:2021eha}. To verify the cross sections and the shapes of the kinematic variable distributions, a comparative analysis was performed on events generated by these packages. One million events per model are used in the analysis.

Additionally, all samples were hadronized in Pythia8~\cite{Bierlich:2022pfr} and run through a detector response simulation with DELPHES~\cite{deFavereau:2013fsa} in order to study the effects of detector smearing on the reconstruction performance. The selection criteria are: 1-2 b-jets and at least 1 non-b jet, in total 2-3 jets with $p_t > 30$ GeV, $\eta < 4.7$, at least 1 lepton with $p_t > 26$ GeV, $\eta < 2.4$ and MET $> 20$ GeV.

The features used for training the models can be conditionally divided into two types: low-level and high-level.
\begin{itemize}
    \item Low-level variables include the components of the 4-momentum for each of the final particles, as well as the missing transverse energy (MET). In cases where particles are absent, the corresponding variables are set to zero.
    \item High-level variables refer to various combinations of low-level variables, justified by physical considerations. These include scalar products between momenta of different particles, reconstructed invariant masses of intermediate particles ($W, t$), and others.
\end{itemize}

The target variables are taken as the three momentum components of the neutrino and the mediator ($p_{\nu_x}, p_{\nu_y}, p_{\nu_z}, p_{\Phi_x}, p_{\Phi_y}, p_{\Phi_z}$). Both input and output variables are standardized to ensure homogeneity of the feature space. The data are divided into training, validation, and test sets in the ratio $0.6:0.2:0.2$.

\subsection{Methodology}
For successful restoration of the required angular correlations between particles, it is necessary to clearly separate the momenta of the neutrino and the DM mediator. This was extensively studied in~\cite{Bunichev:2024}. In~\cite{Abasov:2024nec} an attempt was made to perform analytical calculations for this purpose; however, it did not lead to the desired result. An alternative method is the application of machine learning techniques to restore the momentum components of the neutrino and the DM mediator and to compute the angular variable $cos\theta_{\bar{d} \bar{\ell}}$ based on them.
The entire analysis chain is performed on two datasets: at the parton level and after hadronization and detector smearing. In this work, accuracy is the main priority since any future analyses will be performed offline, where processing speed is not a critical issue.
\subsubsection{Metrics}
\label{sec:metrics}
From a physical point of view, the most optimal criterion for closeness is the relative deviation of the reconstructed distribution from the original. However, since the denominator of such a metric may be near zero, its use is challenging, and therefore the Mean Absolute Error (MAE) is used as the main metric (with the L1 loss function). An analysis was carried out for data consisting only of SM events - a "toy" problem - where the metric value is on the order of $10^{-2}$ and the reconstruction perfectly matches the target data. For the network trained on a 1:1 mixture of SM and DM events, the best MAE value on the test set is 0.3.

For training Normalizing Flows, unlike other architectures, a likelihood function is used. Since its computation requires knowledge of the determinant, constraints are imposed on the network structure to speed up this operation compared to the general case.

Nevertheless, for none of the cases can one specify the desired value of these metrics in the task; the primary criterion will be the closeness of the reconstructed angular variable to the one shown in Fig.~\ref{fig:spincorrs_parton_scal}. Two metrics are used for comparing these distributions during testing:
\begin{itemize}
    \item An analog of MAE for histograms, where the variables are the number of events in each histogram bin; hereafter referred to as hist MAE.
    \item The $\chi^2$ statistic (hereafter $\chi^2$ score), where the expected values are taken as the target number of events in the bins and the observed values as the predicted ones.
\end{itemize}
Using these metrics as loss functions is challenging due to the necessity of computing histograms on a large amount of data.

\subsubsection{Multilayer Perceptron}
\label{sec:mlp}
As a baseline method, a multilayer perceptron (MLP) was chosen, as used in~\cite{Abasov:2024nec}. The model consists of 3 layers with 500 neurons each and a dropout rate of 0.1. The PyTorch library~\cite{Ansel_PyTorch_2_Faster_2024} was used for training the model. This architecture is one of the simplest neural networks yet has proven effective in other tasks related to top quark physics.

\subsubsection{Normalizing Flows and $\nu$-Flows}
\label{sec:nu-flows}
Normalizing Flows is a method based on an invertible transformation of a known probability distribution (for example, the standard normal) into the distribution of the target variables. Since such networks work with the full multidimensional probability density, they can more precisely preserve the dependencies among the target variables. Various approaches to implementing Normalizing Flows are described in detail in~\cite{Kobyzev_2021}.

\begin{figure}[p]
    \centering
    \includegraphics[width=0.5\linewidth]{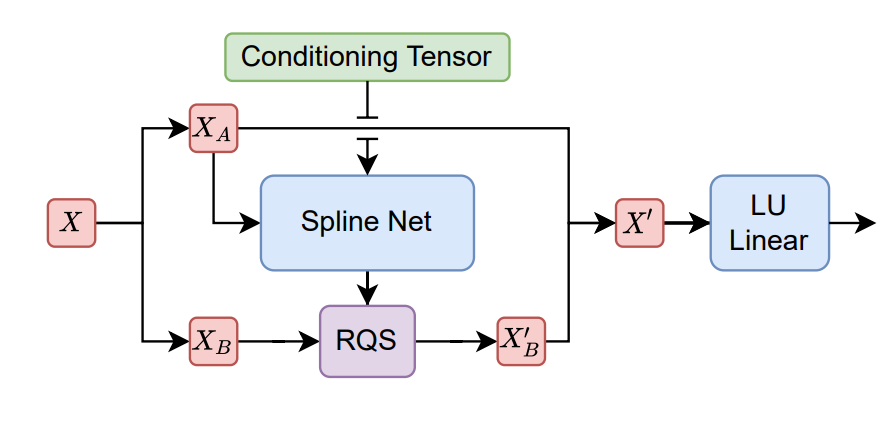}
    \caption{Block of the $\nu$-Flows network~\cite{Leigh:2022lpn}. Here, $X$ represents the target momenta of the neutrino and mediator, the Conditioning Tensor is the output of the encoding fully connected network, and RQS denotes the piecewise rational quadratic splines.}
    \label{fig:nuflows_layer}
\end{figure}

Normalizing Flow architectures have shown good results in similar tasks in $t\bar{t}$ processes, so their use in this task is a natural extention of the current framework. In this work, two models based on Normalizing Flows, implemented using the nflows library~\cite{conor_durkan_2020_4296287}, are studied:
\begin{itemize}
    \item Basic Flows - a "toy" model consisting solely of autoregressive layers. In this model, the contextual variables (outputs of the encoding network) serve as the parameters of normal distributions.
    \item $\nu$-Flows - an architecture following the recommendations in~\cite{Leigh:2022lpn}. It consists of coupling layers that split the target variables into two groups in each layer, one of which remains unchanged while the other undergoes transformations. Piecewise rational quadratic splines are chosen as the transformations. The connection between the two groups of variables is established using a fully connected neural network, to whose input the contextual variables are concatenated. A complete block of the network is shown in Fig.~\ref{fig:nuflows_layer}.
\end{itemize}
For the encoding network in both models, an architecture similar to the baseline model is used but with a reduced number of neurons - 200 instead of 500.
\begin{figure}[p]
    \centering
    \includegraphics[width=0.5\linewidth]{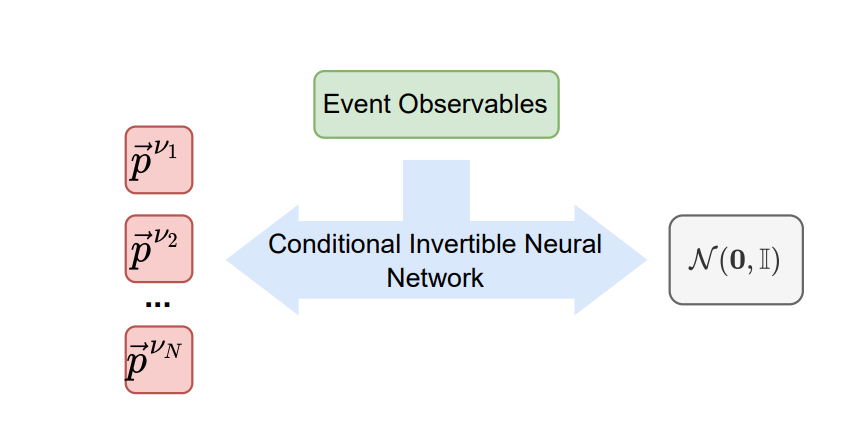}
    \caption{Schematic diagram of Normalizing Flows. Here, Event Observables are variables constructed based on the data from known particles~\cite{Leigh:2022lpn}.}
    \label{fig:nuflows_arc}
\end{figure}

\section{Results}
\label{sec:results}
\begin{figure}[p]
    \centering
    \includegraphics[width=0.32\linewidth]{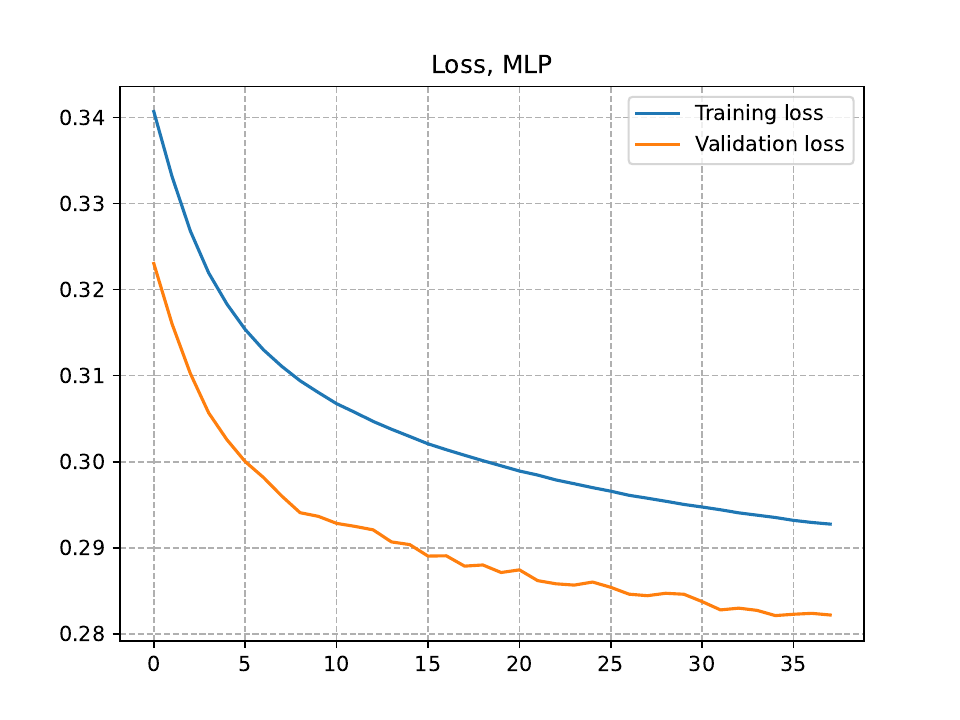}
    \includegraphics[width=0.32\linewidth]{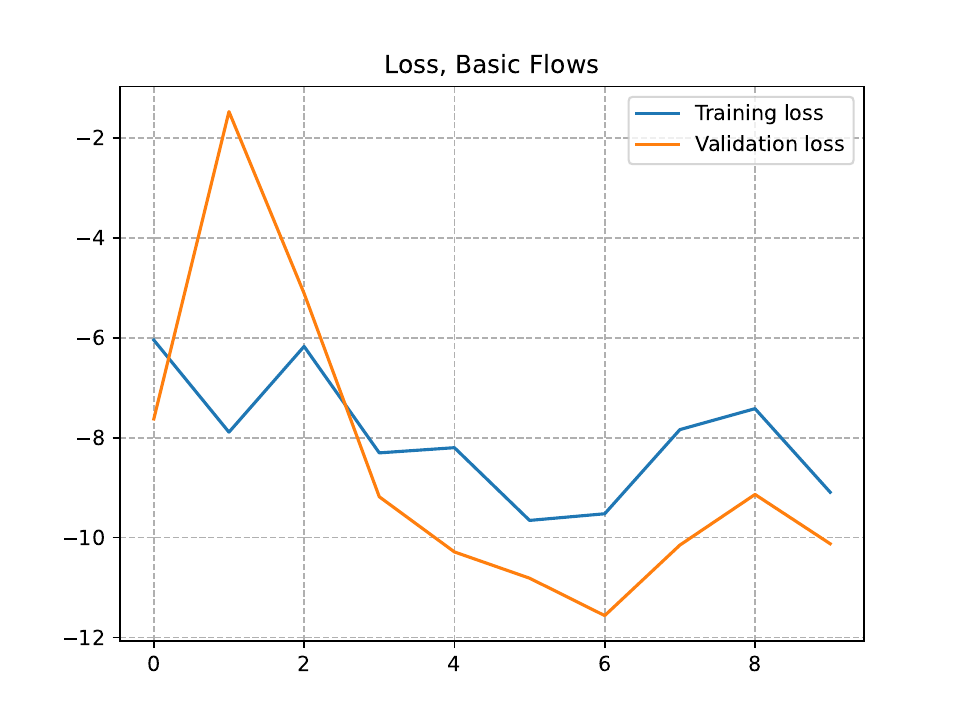}
    \includegraphics[width=0.32\linewidth]{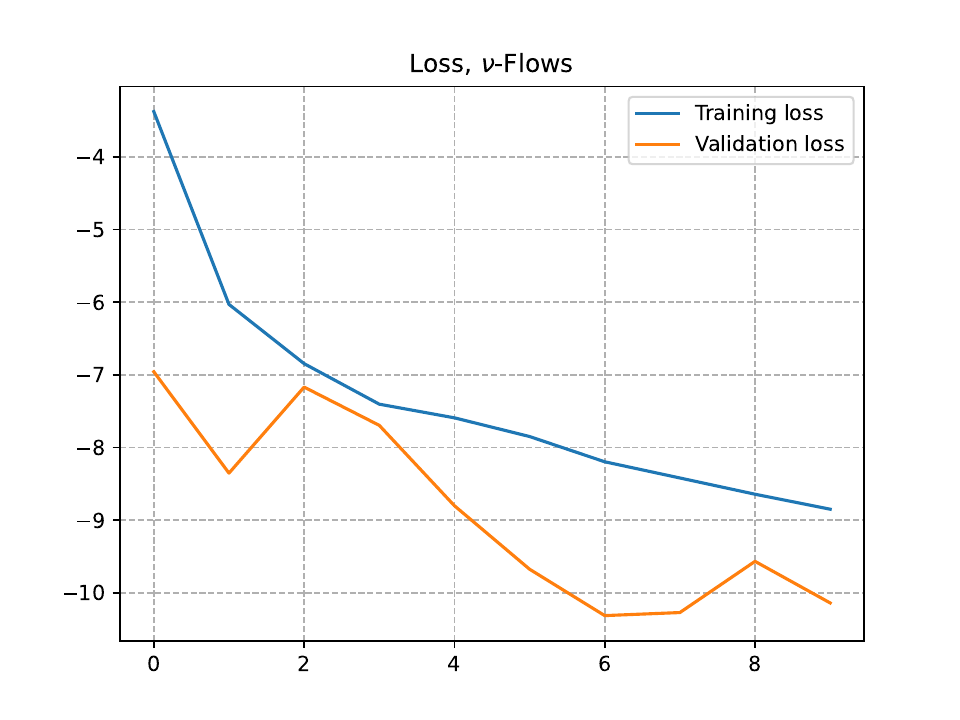}
    \caption{Loss function curves for different models. L1 loss is used for the MLP, and the logarithm of the likelihood function $-2\ln L$ is used for Normalizing Flows.}
    \label{fig:loss}
\end{figure}

This section presents comparisons of the results obtained using various machine learning methods: MLP, Normalizing Flows with autoregressive layers, and Normalizing Flows based on coupling layers. Results are provided for both generator-level data and data after simulating the detector response in DELPHES. The loss function curves for each model are shown in Fig.~\ref{fig:loss}. It is noticeable that the Basic Flows converge worse than all other models; this is manifested in the MAE on the test data tending toward infinity. To address this, the network outputs are clipped to the interval [-10, 10] prior to the inverse transformation. Since Normalizing Flows work with distributions, proper computations require sampling, after which the resulting distribution is aggregated; in this work, the median is computed.

For the $\nu$-Flows network, hyperparameters were tuned using the optuna library, with the $\chi^2$ score on the validation set used as the target metric. The tuning ranges are given in Table~\ref{tab:optuna}.

\begin{table}[ht]
    \centering
    \begin{tabular}{|c|c|c|}
        \hline
        Parameter  & Tuning Range & Chosen Value \\
        \hline
        Encoding vector size (context size)  & [8, 64] & 31 \\
        \hline
        Number of blocks & [3, 7] & 4 \\
        \hline
        Number of neurons in the inner network layer & [32, 256] & 100 \\
        \hline
        Number of inner network layers & [2, 5] & 2 \\
        \hline
    \end{tabular}
    \caption{Optuna parameters for $\nu$-Flows.}
    \label{tab:optuna}
\end{table}

\begin{figure}[p]
    \centering
    \includegraphics[width=\linewidth]{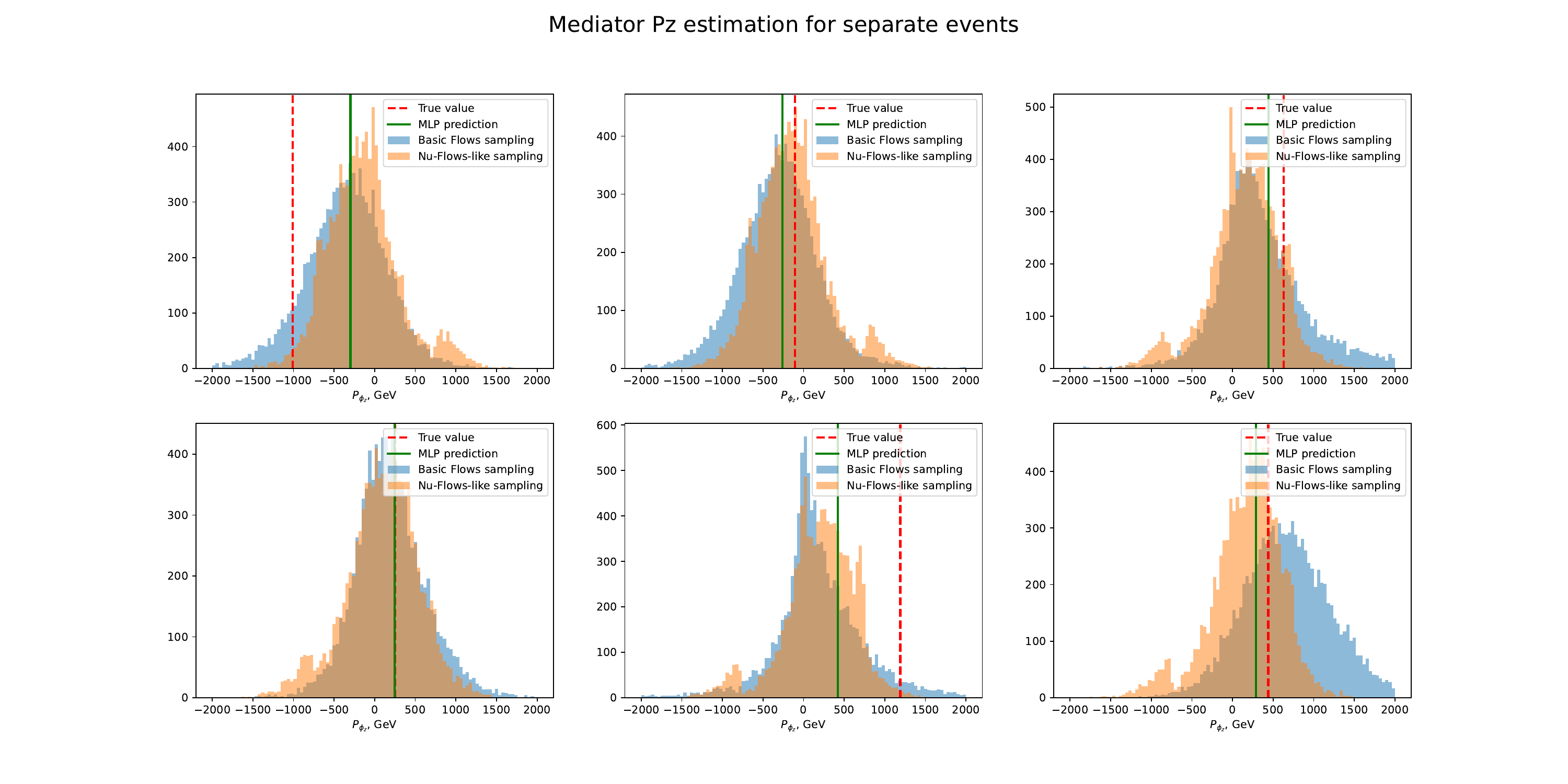}
    \caption{Comparison of the reconstruction of the z-component of the mediator's momentum $P_{z}^{\phi}$ for individual events. For Flow-based networks, 10,000 points are sampled for each event.}
    \label{fig:NF_comparison}
\end{figure}

\begin{figure}[p]
    \centering
    \includegraphics[width=\linewidth]{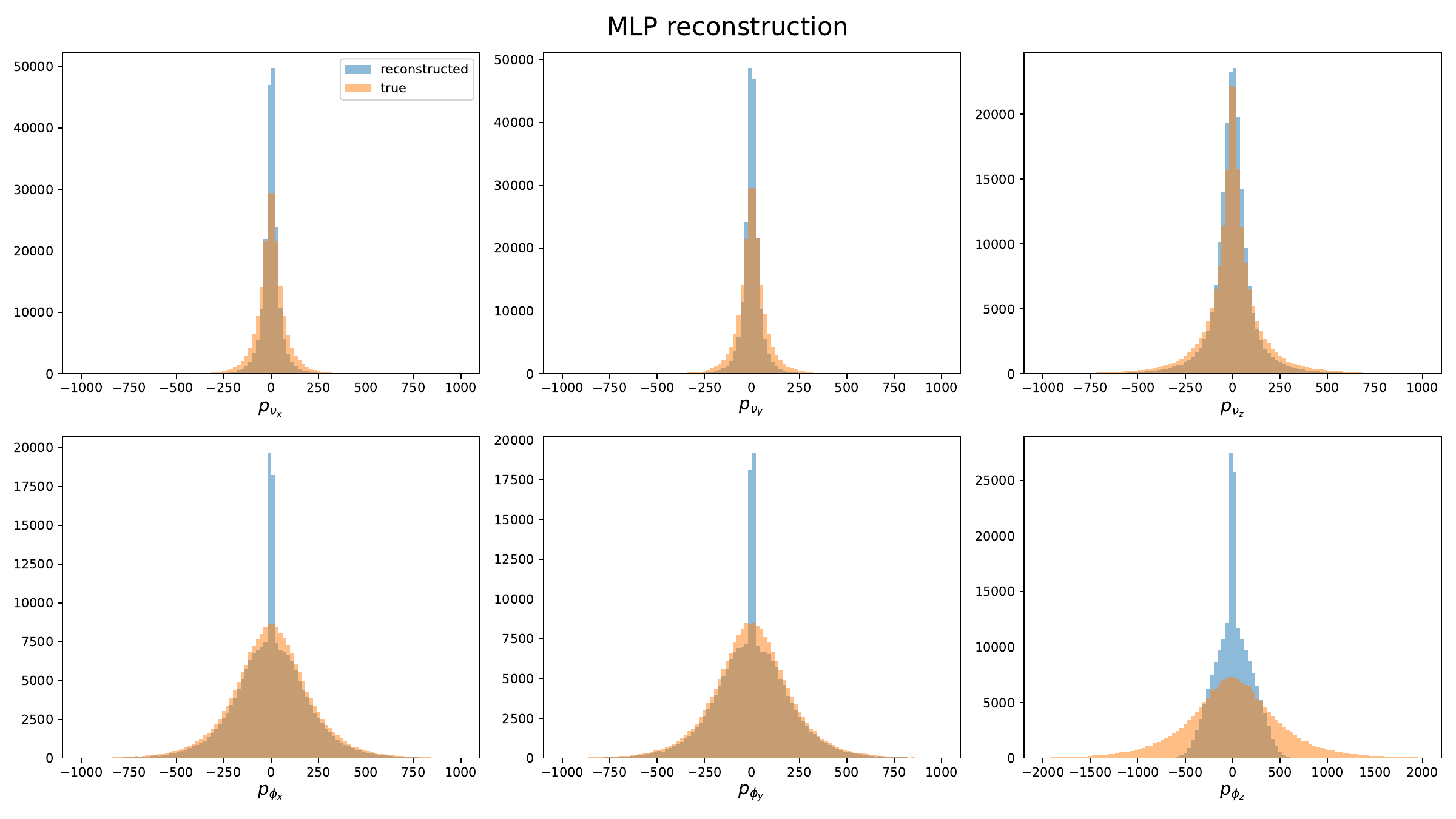}
    \caption{Reconstruction of the momentum components of the neutrino and mediator using the fully connected network.}
    \label{fig:mlp_momentum}
\end{figure}

\begin{figure}[p]
    \centering
    \includegraphics[width=\linewidth]{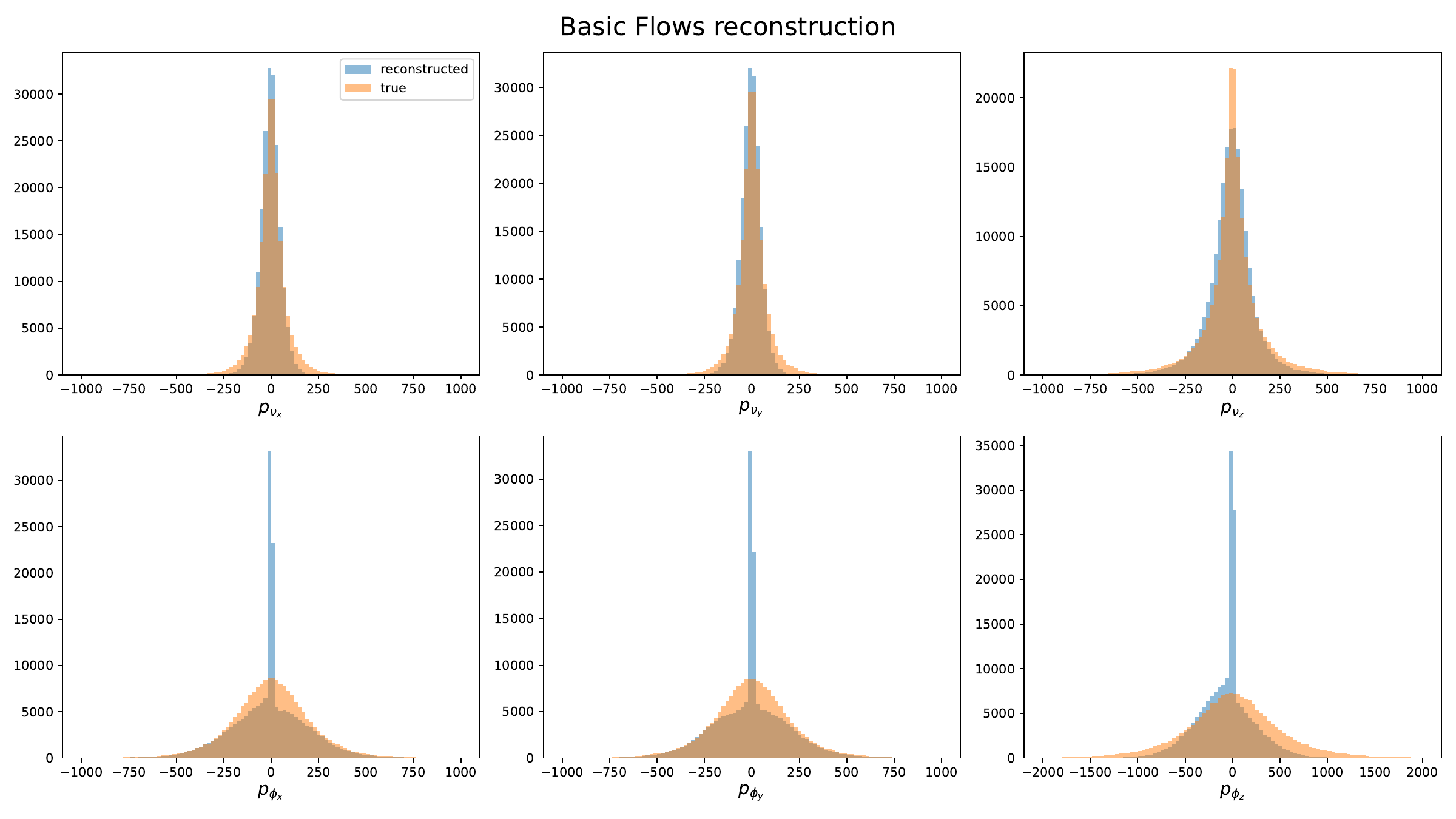}
    \caption{Reconstruction of the momentum components of the neutrino and mediator using Basic Flows. For each event, the median of 10 points is used as the reconstructed value.}
    \label{fig:basic_flows_momentum}
\end{figure}

\begin{figure}[p]
    \centering
    \includegraphics[width=\linewidth]{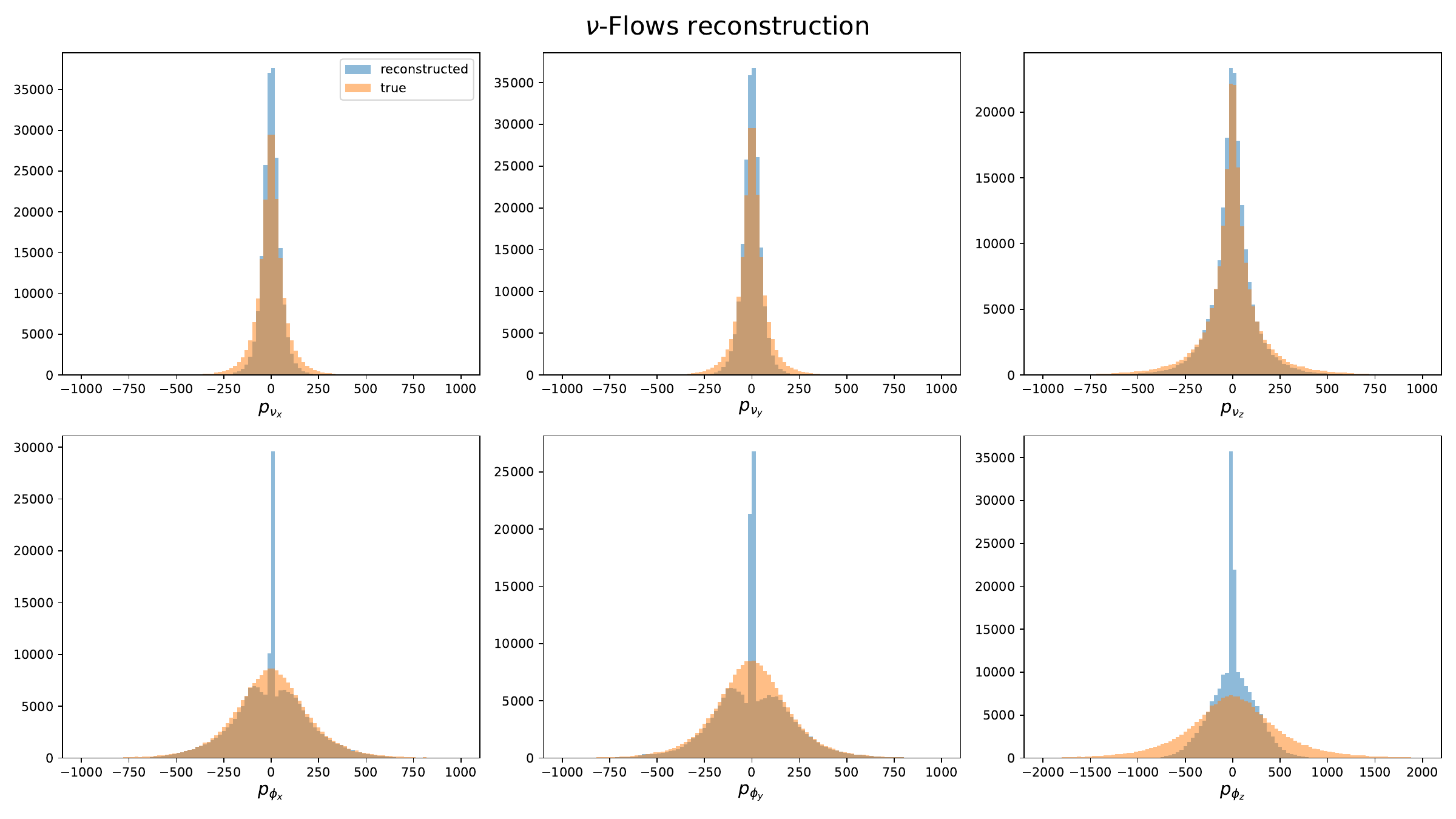}
    \caption{Reconstruction of the momentum components of the neutrino and mediator using $\nu$-Flows. For each event, the median of 5 points is taken as the reconstructed value.}
    \label{fig:nu_flows_momentum}
\end{figure}

The metrics on the test data are given in Table~\ref{tab:metrics}. It can be noted that the fully connected network performs better in terms of MAE but is significantly inferior in the "histogram" metrics, which assess the closeness of the distributions. The reconstruction of the momentum components at the parton level for each network is shown in Fig.~\ref{fig:NF_comparison} for individual events, and in Figs.~\ref{fig:mlp_momentum},~\ref{fig:basic_flows_momentum},~\ref{fig:nu_flows_momentum} for the integrated distributions.

\begin{table}[ht]
    \centering
    \begin{tabular}{||c|c|c|c||}
        \hline
         Model & MAE on momenta & Hist MAE & $\chi^2$ score \\
         \hline
         MLP & 0.5305 & 360.6 & 8985 \\
         \hline
         Basic Flows & 0.6620 & 147.7 & 1557 \\
         \hline
         $\nu$-Flows & 0.6461 & 52.4 & 335 \\
         \hline
    \end{tabular}
    \caption{Comparison of metrics for different architectures at the parton level.}
    \label{tab:metrics}
\end{table}

\begin{figure}[p]
    \centering
    \includegraphics[width=0.32\linewidth]{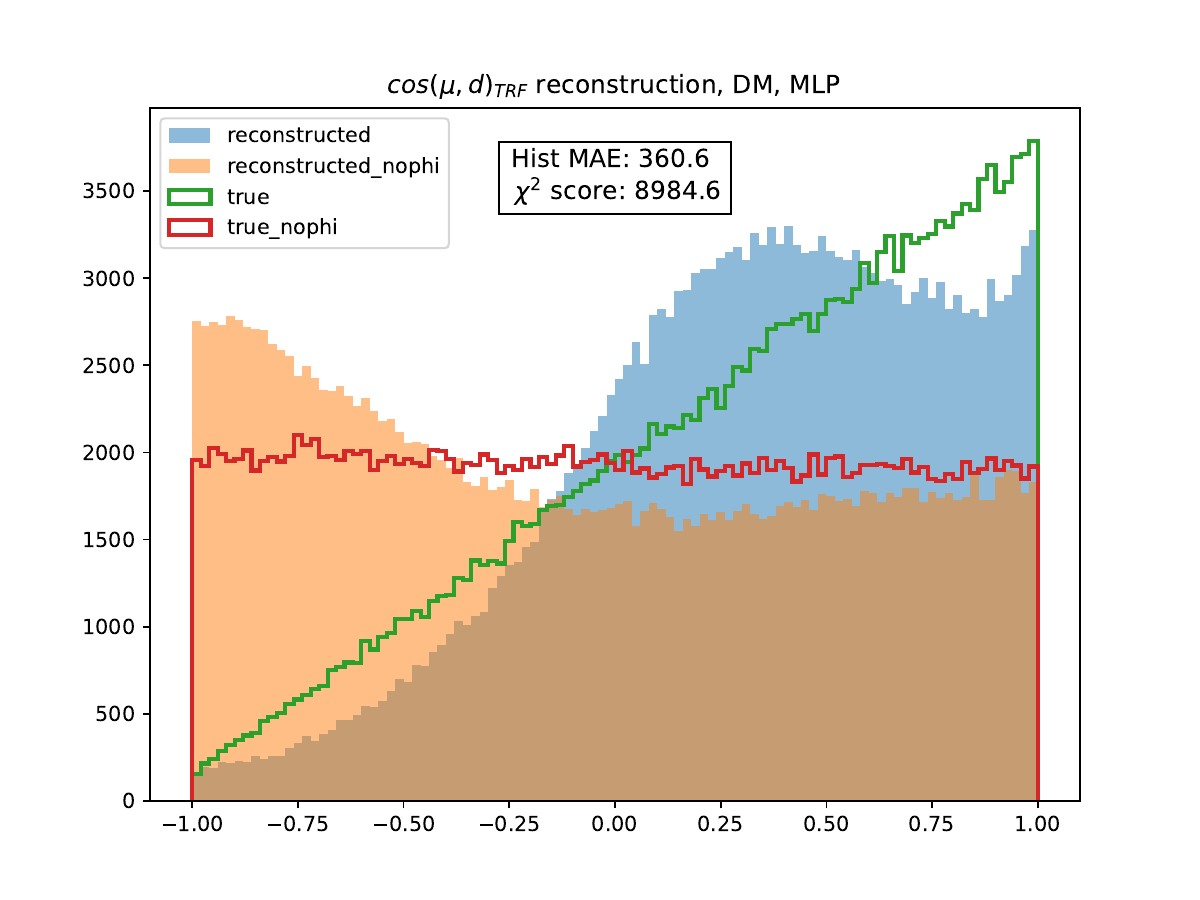}
    \includegraphics[width=0.32\linewidth]{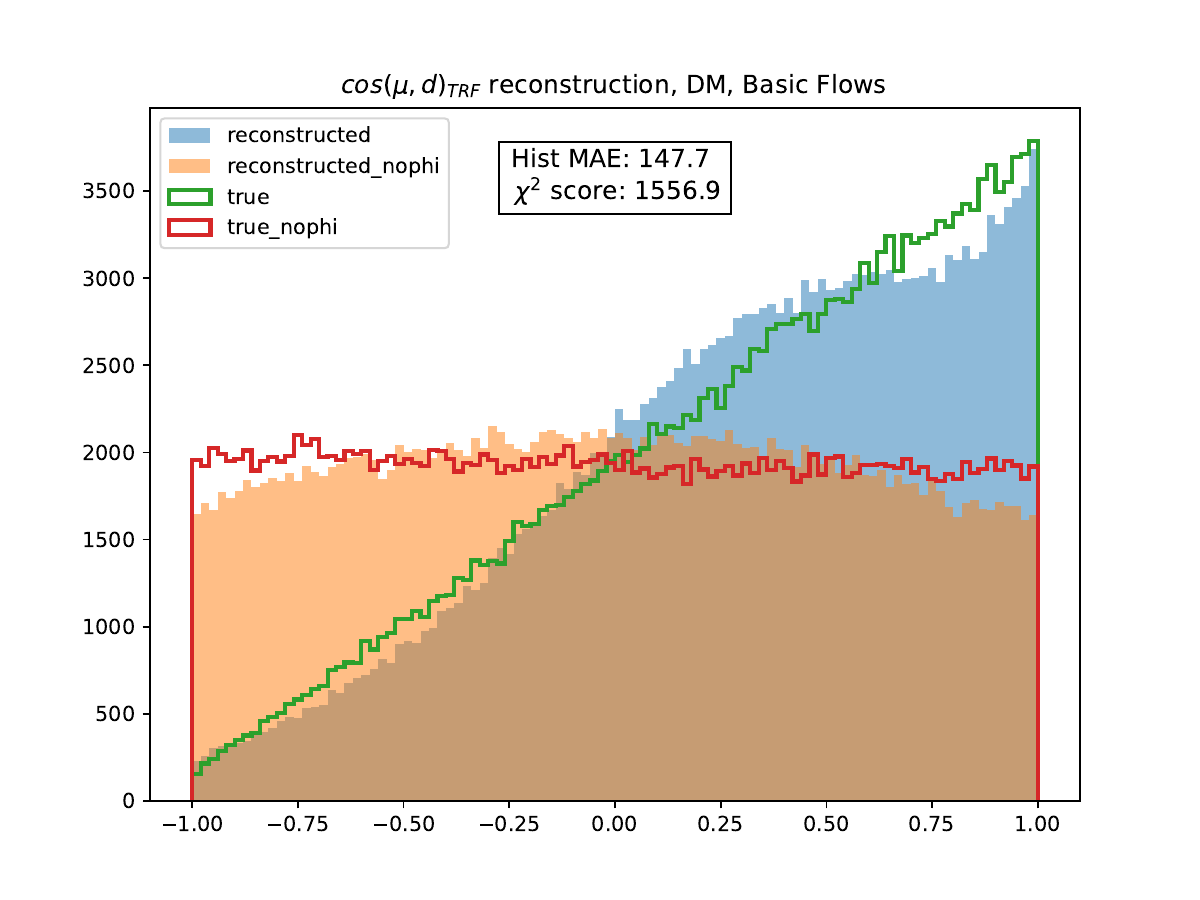}
    \includegraphics[width=0.32\linewidth]{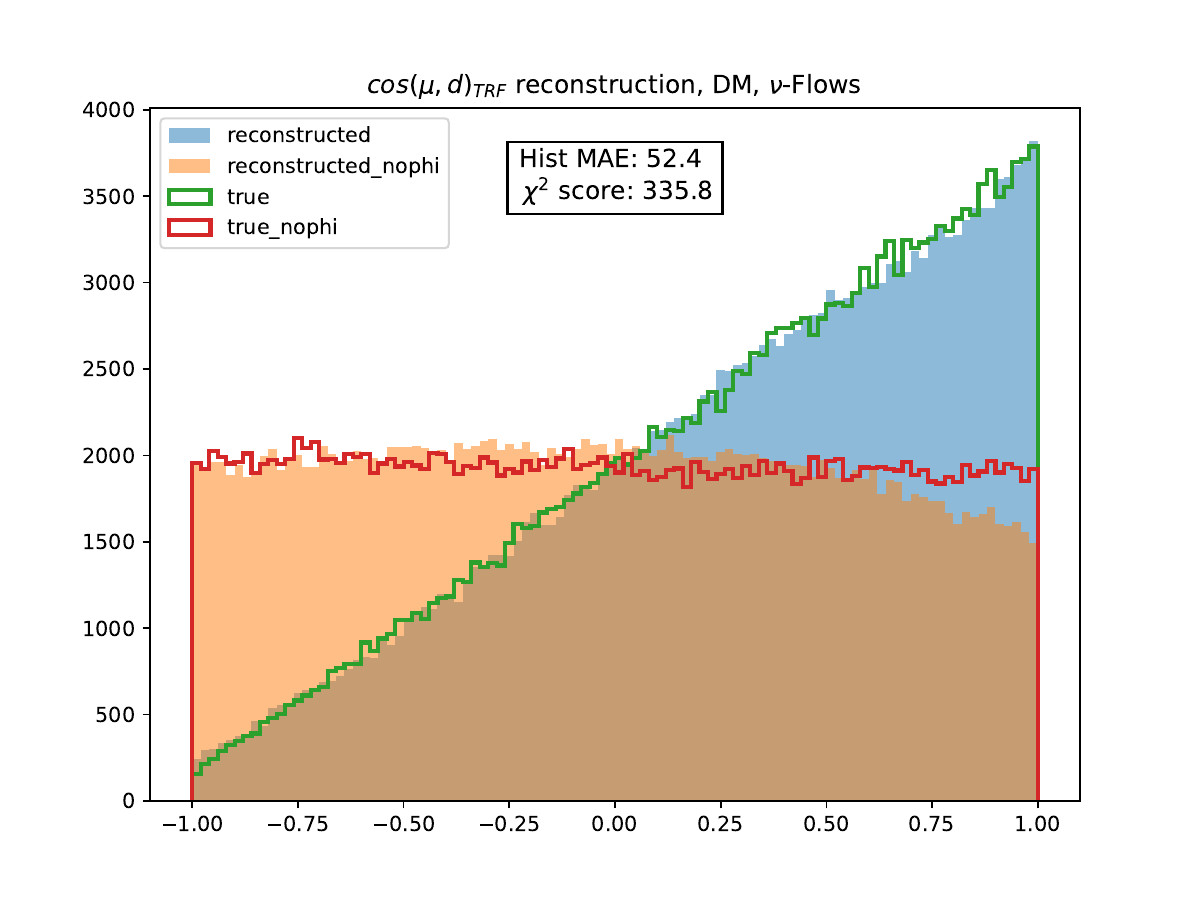}
    \caption{Comparison of the reconstructed and original distributions of the cosine of the angle between the lepton and the down-type quark in the top quark rest frame at the parton level. The postfix "nophi" denotes the top quark rest frame reconstruction in which only the neutrino is used.}
    \label{fig:cos_rec_parton}
\end{figure}

Fig.~\ref{fig:cos_rec_parton} shows a comparison between the reconstructed and original distributions of the cosine of the angle between the lepton and the down-type quark in the top quark rest frame, reconstructed in two ways: one including the mediator in the calculation and one with only the neutrino, which corresponds to the methodology shown in Fig.~\ref{fig:spincorrs_parton_scal}. From both the metric comparison and the visual evaluation of the distributions, it can be concluded that the $\nu$-Flows network reconstructs the target variable most accurately and can be used in real analyses for the search for dark matter.

\begin{figure}[p]
    \centering
    \includegraphics[width=0.32\linewidth]{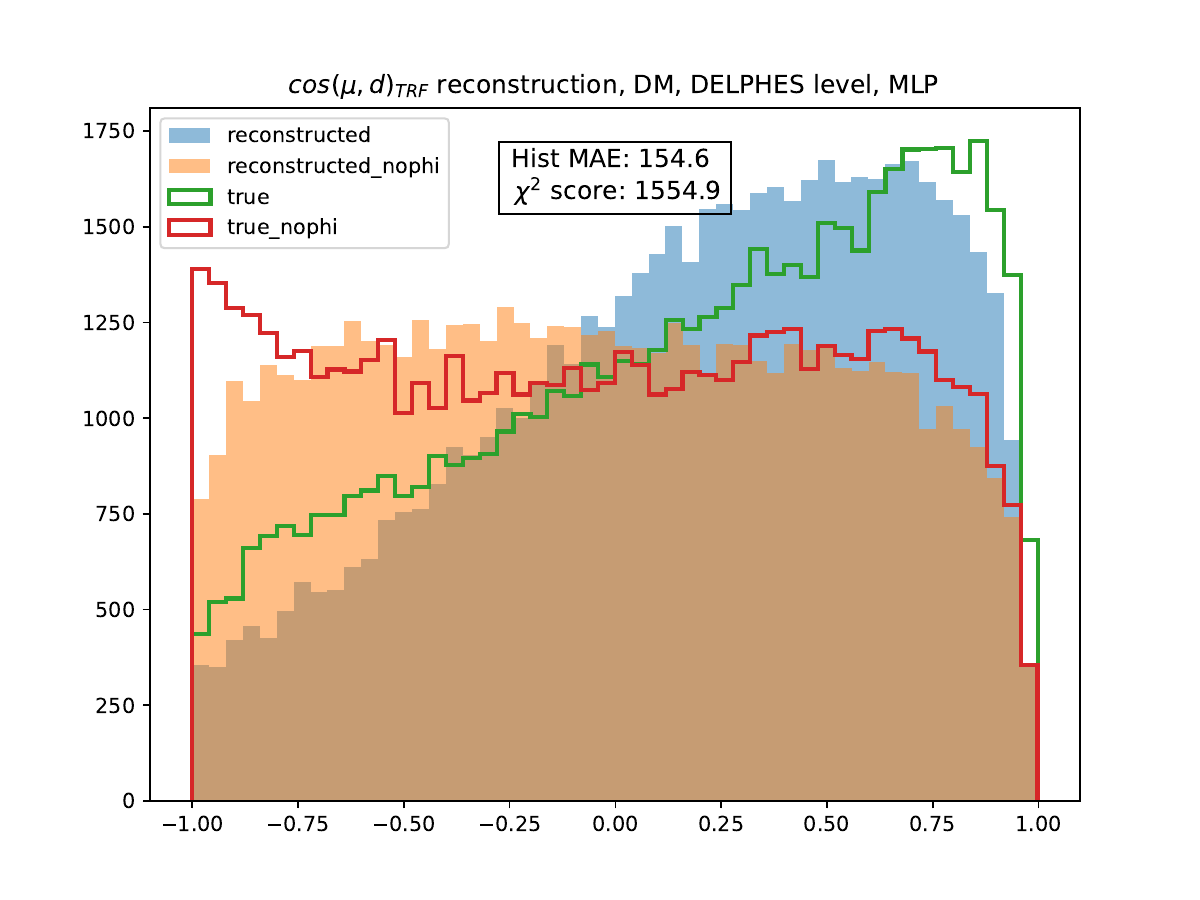}
    \includegraphics[width=0.32\linewidth]{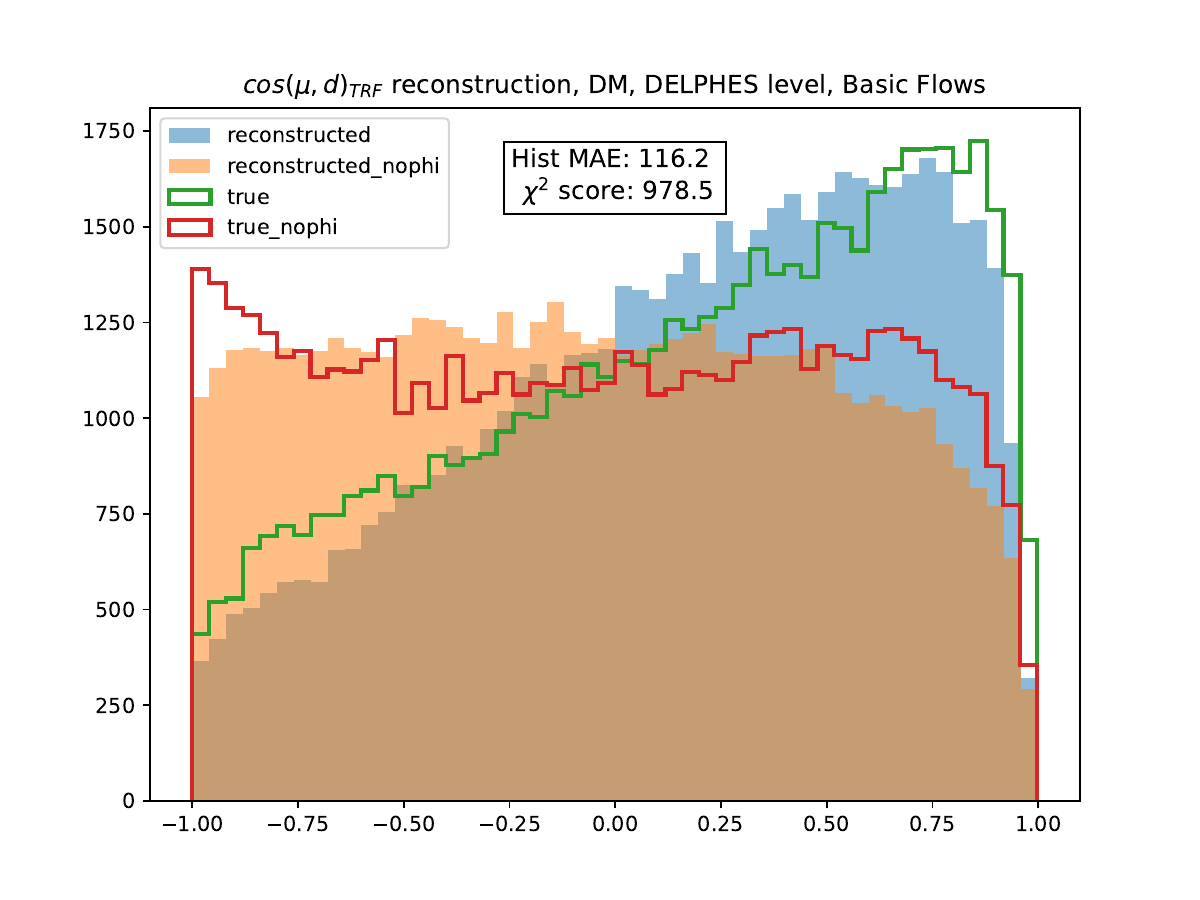}
    \includegraphics[width=0.32\linewidth]{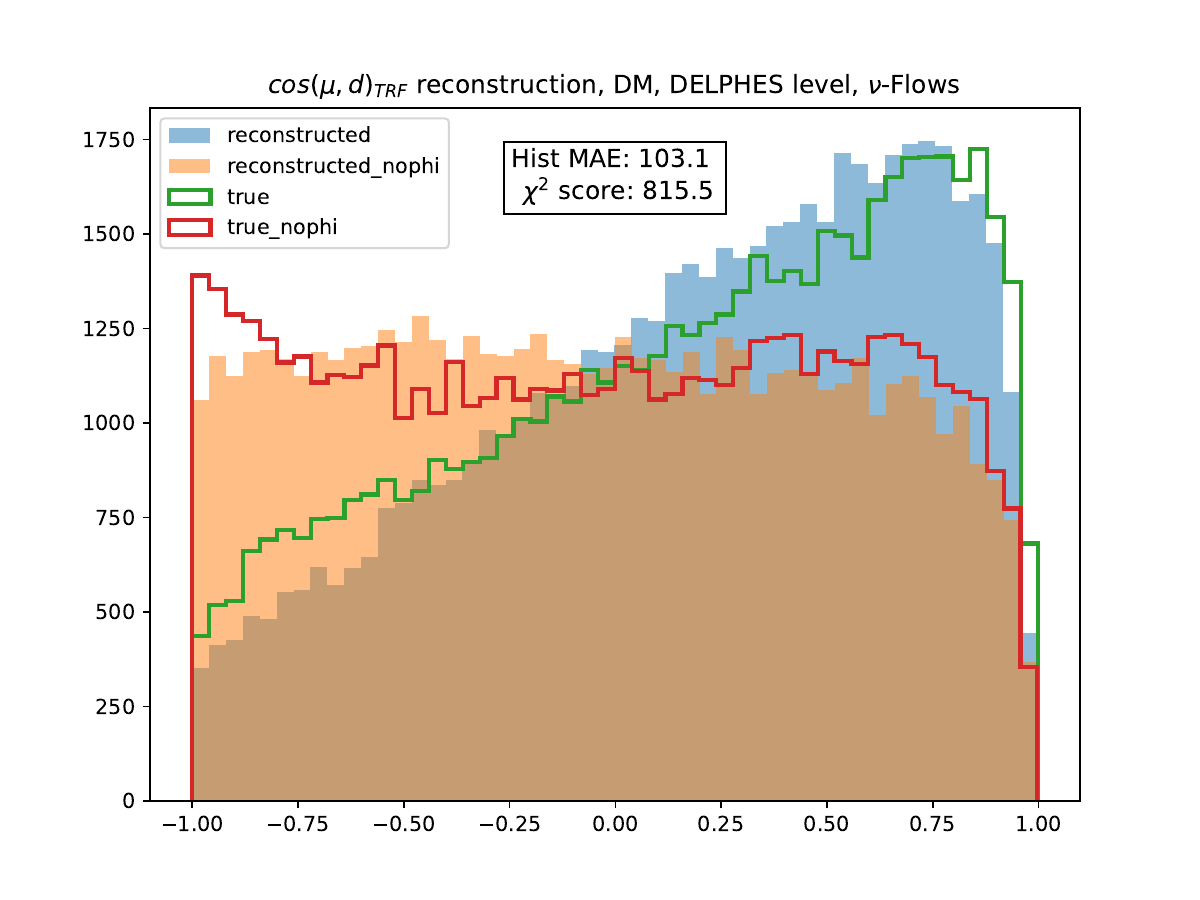}
    \caption{Comparison of the reconstructed and original distributions of the cosine of the angle between the lepton and the down-type quark in the top quark rest frame after detector smearing in DELPHES. The postfix "nophi" denotes the reconstruction of the top quark rest frame using only the neutrino.}
    \label{fig:cos_rec_delphes}
\end{figure}

Detector smearing introduced by DELPHES, as well as tighter selection criteria slightly changes the desired distribution and the network performance slightly drops. However, $\nu$-Flows still outperforms all other methods and provides good reconstruiction quality, the final distributions are presented in Fig.~\ref{fig:cos_rec_delphes}.


\section{Conclusions}
In this work, a new method was proposed for separating the momentum components of the scalar dark matter mediator and the neutrino in single top quark production based on the Normalizing Flows architecture. This approach outperforms multilayered perceptrons, significantly improves the accuracy of reconstructing the top quark's angular correlations at both the parton level and after simulation of the detector response and can be directly applied to collider data.
\section*{Code}
All of the results described, as well as the model architectures, are available on GitHub~\cite{emil_abasov_2025_15242257}.
\section*{Acknowledgments}
Study in sections 1, 2.1, 2.2.1, 3 was conducted within the scientific program of the Russian National Center for Physics and Mathematics, Section 5 "Particle Physics and Cosmology".

The research in sections 2.2.2 and 2.2.3 was conducted by E.Abasov with support from Non-commercial Foundation for the Advancement of Science and Education INTELLECT.

\clearpage

\clearpage
\bibliographystyle{maik}
\bibliography{main}

\end{document}